\documentclass[prd,showpacs,floatfix]{revtex4}
\usepackage{graphicx}

\def\IP{I\!\!P}

\begin{document}
\title{Diffractive effects in spin-flip $pp$ amplitudes and 
predictions for relativistic energies}
\author{A. F. Martini}
 \email{martini@ifi.unicamp.br}
\affiliation{Instituto de F\'{\i}sica Gleb Wataghin, Universidade 
Estadual de Campinas,\\
Unicamp, 13083-970 Campinas, S\~ao Paulo, 
Brazil}

\author{E. Predazzi}
 \email{predazzi@to.infn.it} 
\affiliation{Universit\'a di Torino and INFN, Sezione di Torino,\\
Via Pietro Giuria, 1, 10125 Torino, Italy}


\begin{abstract}
We analyze the diffractive (Pomeron) contribution to $pp$ 
spin-flip amplitude and discuss the possible scenarios for energies 
available at the Relativistic Heavy-Ion Collider (RHIC). In particular,we 
show that RHIC data will be instrumental in assessing the real 
contribution of diffraction to spin amplitudes. 
\end{abstract}


\centerline{PHYSICAL REVIEW D {\bf 66}, 034029 (2002)}

\maketitle

\section{\label{sec:intro}Introduction}

The study of diffraction on high-energy scattering using Regge formalism 
is a well-known subject \cite{predazzi98}. Until the coming in operation 
of Relativistic Heavy-Ion Collider (RHIC), the highest energy $pp$ 
data available from accelerators were 
those of the Intersecting Storage Ring (ISR) at CERN. Those experiments, 
however, did not measure the polarization of the projectiles. Since the 
differential cross section ($d\sigma/dt$) at ISR energies is usually 
assumed to be dominated by the spin-non-flip amplitude, many of the models 
describing the diffraction in $pp$ scattering do not take into account the 
contribution from spin-flip amplitudes 
\cite{desgrolardII00,desgrolard94,donnachie84}. Others do 
\cite{bourrely79,goloskokov91,buttimore99,akchurin99} 
but in this case they use data at lower energies \cite{kline80}. 
However, the study of diffraction in spin-flip amplitudes began many 
years before that when one of us \cite{predazzi67} noticed that 
polarization data at  $\pi p$ scattering suggested a diffractive 
contribution in the spin-flip  $\pi p$ amplitude at high energy which 
becomes evident when the kinematical zero is removed. The spin-flip 
amplitude without the kinematical zero is named ``reduced'' 
and manifests itself as a peak in the forward direction which does not 
appear to vanish as the energy increases. 
In Ref. \cite{predazzi67} it was explicitly noticed that, once the 
kinematical zero is removed, all partial waves act coherently in the 
small angle domain as it is typical of diffractive events 
\cite{good60}. The following statement was made later \cite{hinotani79}: 
``{ \it the residual spin-flip amplitudes behave very much like 
spin-non-flip amplitudes at high energies and exhibit a pronounced 
forward peak which is largely independent of the particular elastic 
reaction chosen}''. The same conclusion was obtained in the 
analysis of $pp$ data some years later \cite{hinotani79} but in 
the case of $pp$ scattering the situation is complicated by the 
existence of five independent helicity amplitudes, named $\phi_j$, 
with $j=1,\ldots , 5$. Of those, $\phi_1$ and $\phi_3$ are 
spin-non-flip amplitudes, $\phi_2$ and $\phi_4$ are double 
spin-flip amplitudes, and $\phi_5$ is a single spin-flip amplitude.

In this work we are interested in high energy and not too high 
$t$ so that we can concentrate the analysis on the main aspects 
of the spin-flip amplitude for $pp$ scattering in the diffractive 
region. In Ref. \cite{buttimore99} the magnitudes of $\phi_2$ and 
$\phi_4$ with respect to the spin-non-flip amplitude were analyzed 
and reasonable arguments were given concerning the linear
dependence on $t$ for $\phi_2$ ($\phi_2\propto t$ when 
$t\to 0$) \footnote{A similar dependence was also utilized in another 
work where the impact parameter  space was 
used \protect\cite{goloskokov91}.}, while in the case of $\phi_4$ 
the relation $\phi_4\propto t$ is consequence of angular momentum 
conservation.

So, the general form of the polarization $P$ for $pp$ scattering 
\begin{equation}
P=2{{\rm Im}[(\phi_1+\phi_2+\phi_3-\phi_4)\phi_5^{\ast}]\over 
[|\phi_1|^2+|\phi_2|^2+|\phi_3|^2+|\phi_4|^2+4|\phi_5|^2]}
\end{equation}
can be simplified assuming that $\phi_2$ and $\phi_4$ are small 
compared to the others amplitudes in the $t$ region under interest. 
Using the definitions
\begin{equation}
\phi_1={g(s,t)\over 2},\quad \phi_5={h(s,t)\over 2},
\label{effspamp}
\end{equation}
where $g(s,t)$ and $h(s,t)$ will be considered 
{\it effective} spin-non-flip and spin-flip amplitudes, respectively, 
and assuming $\phi_1=\phi_3$ the polarization can be rewritten as 
\begin{equation}
P=2{{\rm Im}[g(s,t)h^{\ast}(s,t)]\over |g(s,t)|^2+2|h(s,t)|^2} .
\end{equation}

Today, a rather considerable amount of data at higher energies has been 
gathered \cite{kline80} in the relatively small angle domain and new 
perspectives are being opened by the coming in operation of the 
Relativistic Heavy Ion Collider (RHIC), the ideal machine to study 
polarization in high energy collision processes \cite{guryn00}.

In addition, our phenomenological information on the spin-non-flip 
amplitude is today much more complete and this can be used to reduce 
the uncertainties in the analysis. 
Using an explicit parametrization for the $pp$ spin-non-flip amplitude 
\cite{desgrolard00} whose parameters have been calculated against all 
high energy $pp$ and $\overline{p}p$ data (except polarization), we 
analyze the structure of the reduced spin-flip contribution, where 
the kinematical zero is removed with the factor $\sqrt{-t}$. The 
following conclusions  are reached in this analysis.

(a) The (reduced) spin-flip amplitude exhibits the typical peak in 
the forward direction which characterizes diffractive amplitudes but the 
reduced spin-flip amplitude is less than 10\% of the imaginary part of 
the spin-non-flip amplitude.
(b) The data fitting with the same energy dependence in $g(s,t)$ 
and $h(s,t)$ seems the best choice; a zero is automatically produced in 
the polarization; this moves with energy towards $t=0$ as a consequence of 
the analogous shift of the dip in $d\sigma/dt$ induced by the zero of the 
spin-non-flip amplitude.
(c) The magnitude of the polarization decreases as the 
energy increases but the extrapolation to 500 GeV 
predicts a non-negligible contribution if the same Pomeron 
trajectory for both spin-flip and spin-non-flip amplitudes is used.

In the next section we present the expressions of the amplitudes and 
explain them. In Sec. \ref{sec:fit} we show the fit of the data 
and the modifications on the spin-flip amplitudes. We also consider a 
similar (nondiffractive) analysis of the data. Although the fit to the 
lower energy data is essentially the same, the two analysis predict a very 
different behavior at increasing energies. As we will stress, RHIC data 
should provide a reasonably clear cut answer to the question of whether or 
not diffraction contributes to the spin-flip amplitude. In Sec. 
\ref{sec:concl} we present the conclusions.

\section{\label{sec:defin}Definition of the amplitudes}

The {\it effective} $pp$ spin-non-flip amplitude will be written as
\begin{equation}
g(s,t)=a^{nf}(s,t)=a_{+}(s,t)-a_{-}(s,t)
\end{equation}
with
\begin{equation}
a_{+}(s,t)=a_{\IP}(s,t)+a_f(s,t) \quad {\rm and} \quad
a_{-}(s,t)=a_{O}(s,t)+a_{\omega}(s,t),
\end{equation}
where $a_{\IP}(s,t)$ and $a_{O}(s,t)$ are the Pomeron and Odderon 
amplitudes, respectively, and $a_f(s,t)$ [$a_{\omega}(s,t)$] is the 
even (odd) secondary Reggeon \footnote{Actually, $a_f$ embodies 
both $f$ and $\rho$ contributions (and $a_{\omega}$ both $\omega$ and 
$a_2$).}. These different amplitudes are taken directly from Ref. 
\cite{desgrolard00} and their explicit forms are given in the Appendix 
\ref{appe:spin} together with the values of their parameters.

To write the effective spin-flip amplitude $h(s,t)$: (i) first, we 
neglect the spin contribution of secondary Reggeons to check whether 
diffraction could be the dominant phenomenon for the spin-flip 
amplitude at small $|t|$ at the available energies; (ii) second, 
we notice that the combination of the exchange of two and three 
gluon ladders induce {\it CP}-even (and -odd) contributions to 
the spin-flip amplitude that affects Pomeron exchange. 
We have, therefore, to anticipate that the diffractive (Pomeron) 
contribution to the spin-flip amplitude may have a slightly modified 
structure from the usual one. Differently stated, to take into account 
three gluon ladders, we do not take a strict Regge pole parametrization 
for the Pomeron spin-flip term and we allow a real part contribution to be 
present. 
Thus, in the very small $|t|$ domain, we write
\begin{eqnarray}
h(s,t)&=&a^{sf}(s,t) = (i\gamma_1+\delta_1){\sqrt{-t}\over m_p}\;
\tilde{s}^{\alpha^{sf}(t)}e^{\beta_1 t}\Theta(0.5-|t|)   
\nonumber\\
&+& (i\gamma_2+\delta_2){\sqrt{-t}\over m_p}\;
\tilde{s}^{\alpha^{sf}(t)}e^{\beta_2 t}\Theta(|t|-0.5),
\label{spinampl2}
\end{eqnarray}
where the mass of the proton $m_p$ is used to make the parameters 
dimensionless. In Eq. (\ref{spinampl2}) 
$\tilde{s}=(s/s_0)e^{-i\pi/2}$, $\Theta$ is the step function 
and we assume $s_0=1\;{\rm GeV^2}$ as in Ref. \cite{desgrolard00}; the 
superscript $sf$ (for ``spin-flip'') allows us to check if 
the $\IP$ trajectory can be different for spin-flip and 
spin-non-flip amplitudes. While the complex phase in Eq. (\ref{spinampl2}) 
is used to take into account in a phenomenological way the mixing of 
{\it CP} even and odd contributions cited above, the step function 
is used to account in the most economical way for the fact that 
the fitting procedure requires a change in the slope of $h(s,t)$  
(similar to what happens in $d\sigma/dt$) to have a good description 
of the polarization data. The precise point where the slope 
changes, however, is rather $t$ insensitive 
and could be taken at almost any value between 0.2 and 0.7 
${\rm GeV}^2$ (in $|t|$). We do not have an explanation for 
this effect nor for the insensitivity in the choice of the 
point where the slope changes but it proves that our 
parametrization is quite stable. 
To start with, we take the spin-flip Pomeron trajectory 
$\alpha^{sf}(t)$ to be exactly the same as derived for 
the spin-non-flip amplitude 
\begin{equation}
\alpha^{sf}(t)=\alpha_{\IP}(t)=\alpha_{\IP}(0)+\alpha_{\IP}'t ,
\label{alfsf}
\end{equation}
where $\alpha_{\IP}(0)$ and $\alpha_{\IP}'$ are given in the Appendix 
\ref{appe:spin}.

\section{\label{sec:fit}Fit to the existing $pp$ polarization data}

The differential cross section $d\sigma/dt$ for $pp$ scattering is 
available at energies up to 63 GeV in the c.m. system but 
polarization data are available only at lower energies. We utilized 
the data at $\sqrt{s}=$ 13.8, 16.8, and 23.8 GeV (a total of 64 
points) in the fit and we checked the quality of the result 
describing the polarization at 19.4 GeV. The 
parameters obtained in the fit are shown in Table \ref{tab6param} with the 
best $\chi^2$ per degree of freedom. In principle (and in practice), the 
inclusion of the spin-flip amplitude in the game would require refitting 
all the parameters to reproduce the angular distributions as well. We 
will check, however, that the fit of the polarization data {\it without} 
performing a new fit to the angular distribution does not modify the 
quality of the fit to the latter while making much more direct the 
analysis of the polarization data

\begin{figure}
\includegraphics[width=9cm,height=7cm]{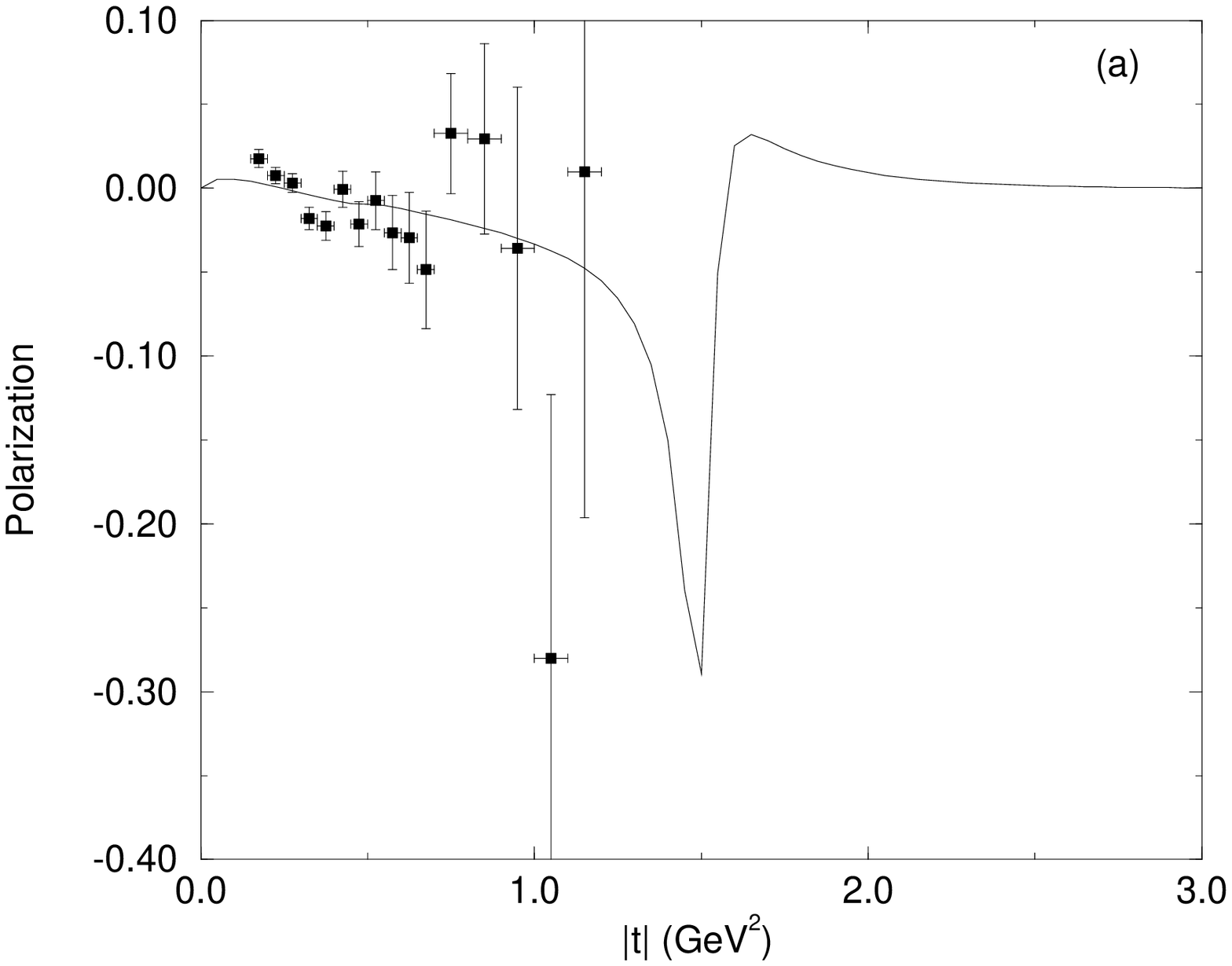}
\includegraphics[width=9cm,height=7cm]{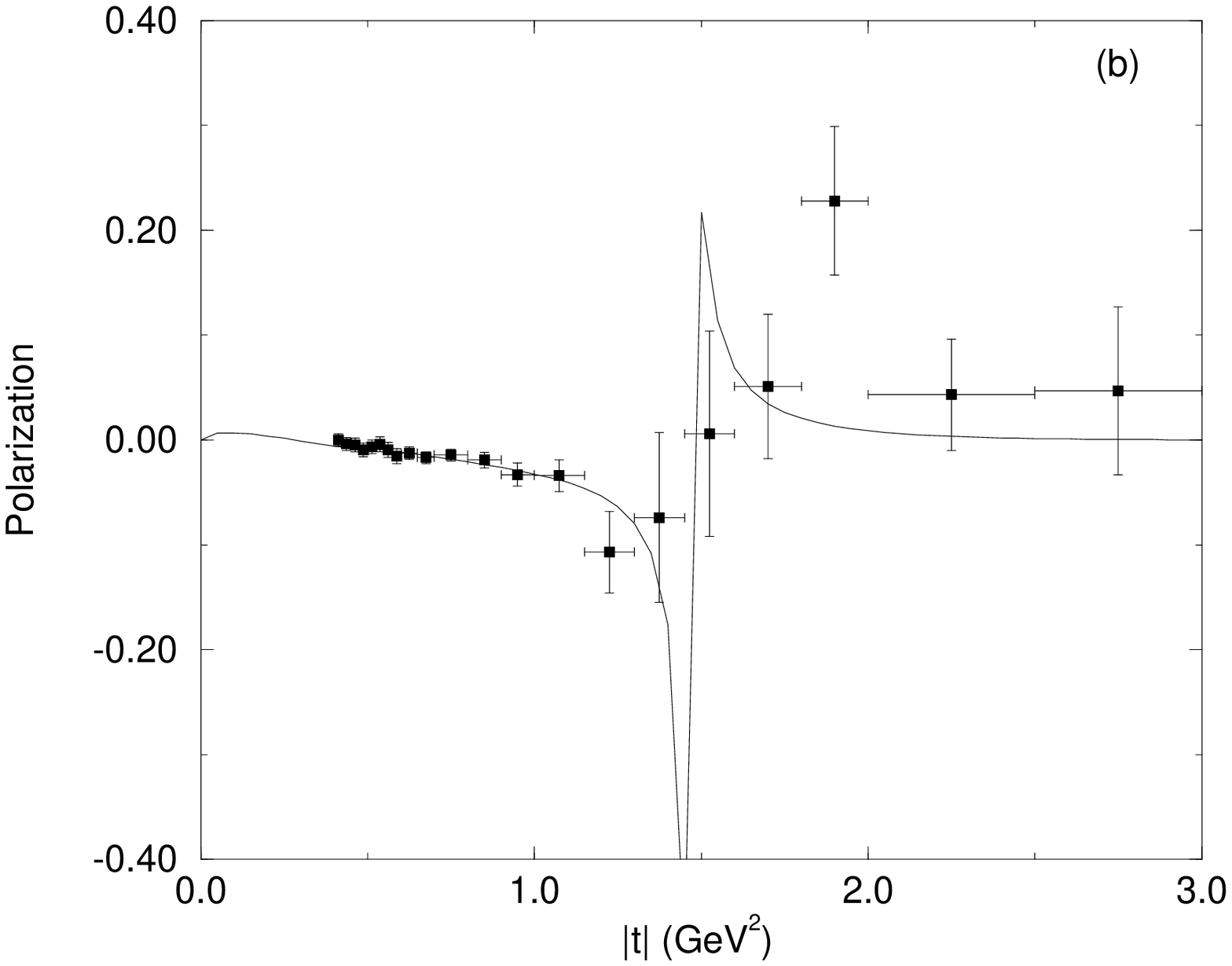}
\includegraphics[width=9cm,height=7cm]{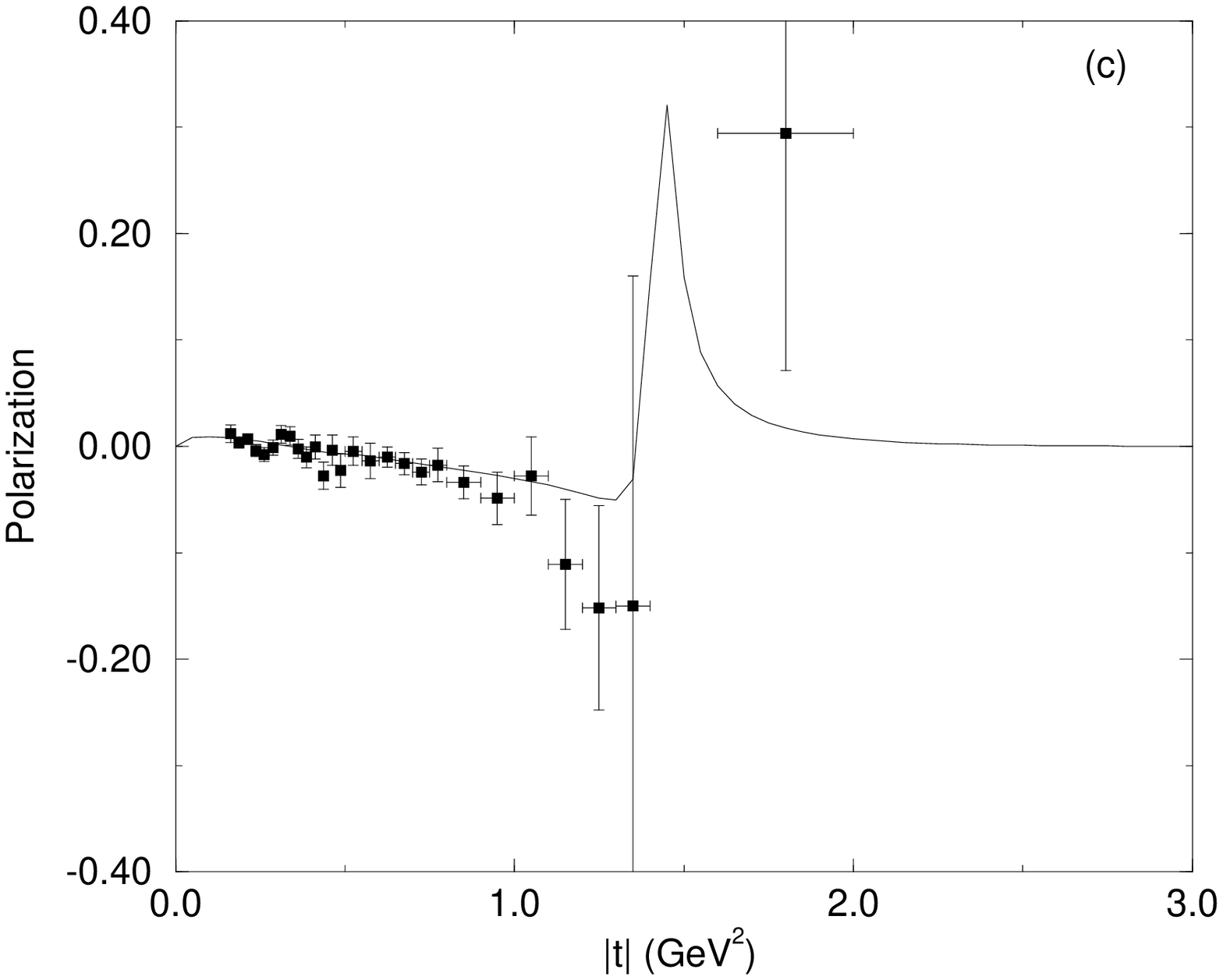}
\caption{\label{figpol6param}Results from fitting polarization data 
at (a) 13.8 GeV, (b) 16.8 GeV, and (c) 23.8 GeV (see Table
\protect\ref{tab6param}).}
\end{figure}

\begin{figure}
\includegraphics[width=10cm,height=8cm]{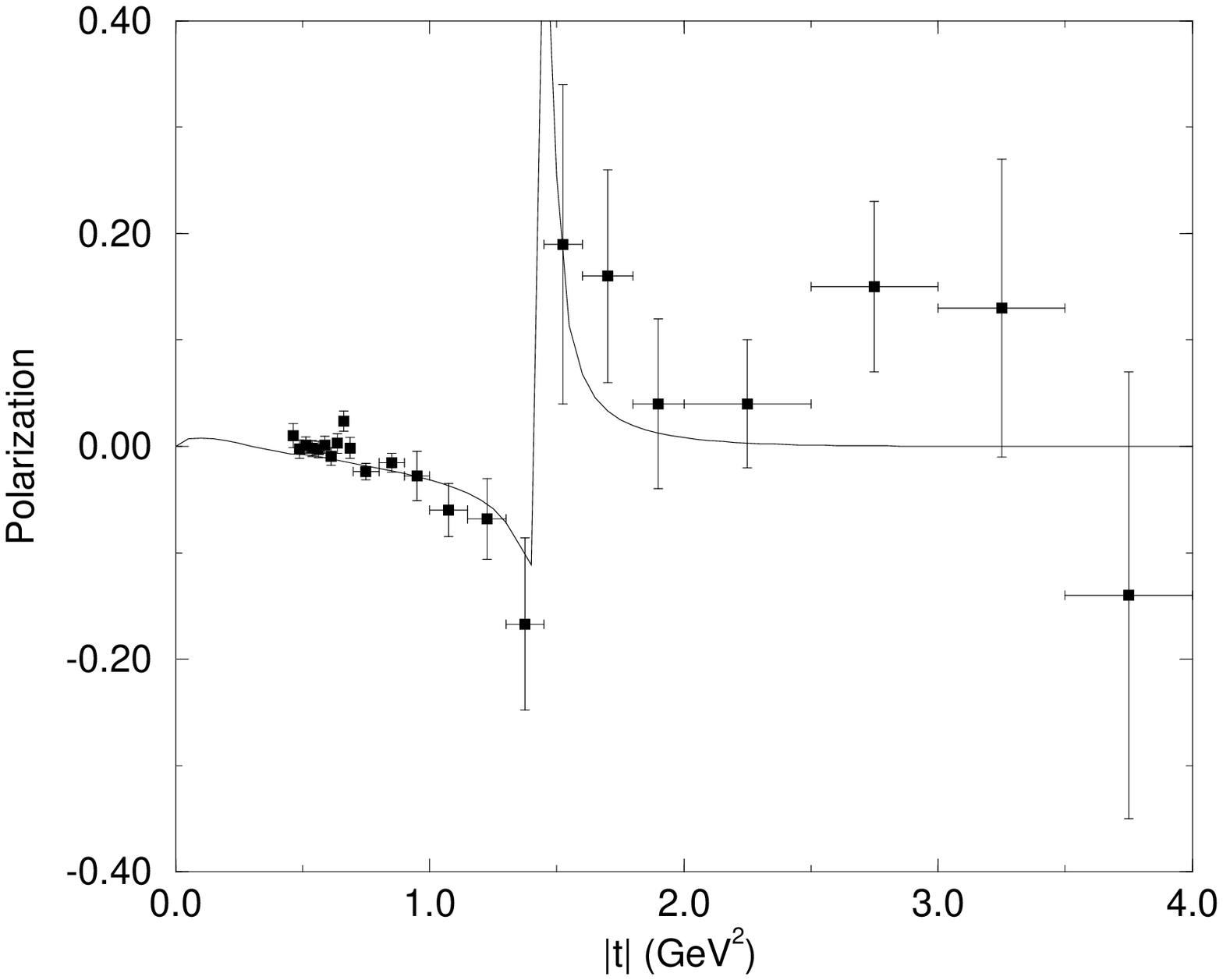}
\caption{\label{figpol19pt4}The prediction for the polarization at 
19.4 GeV (not used in the fit) compared with the experimental data at 
that energy (parameters from Table \protect\ref{tab6param}).}
\end{figure}

\begin{figure}
\includegraphics[width=12cm,height=10cm]{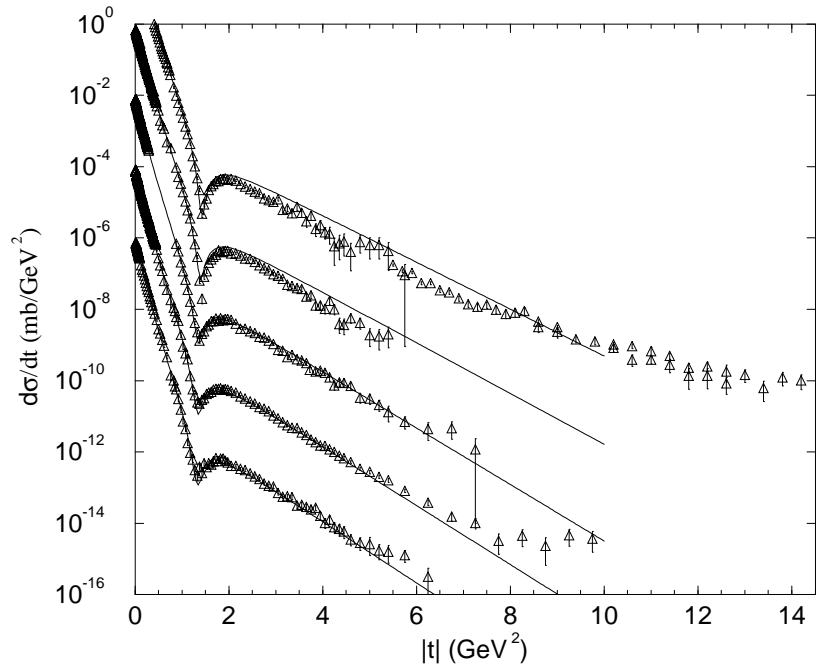}
\caption{\label{figdsdt}The differential cross section obtained in this 
work taking into account the spin-flip amplitude, Eq. 
(\protect\ref{spinampl2}). The highest set of data
corresponds to 23.5 and 27.4 GeV grouped together. The other sets
(multiplied by powers of $10^{-2}$) are 30.5, 44.6, 52.8, and 62 GeV.}
\end{figure}

Figure \ref{figpol6param} presents the set of polarization data used 
in the fit together with our reconstruction. Figure \ref{figpol19pt4} 
shows the polarization at $\sqrt{s}=19.4$ GeV with the prediction of 
our model since this set was not used in the fit. In Fig. \ref{figdsdt} we 
show $d\sigma/dt$ at various energies as described by adding the (squared) 
spin-flip amplitude [see Eq. (\ref{scdif}) in the Appendix 
\ref{appe:spin}] to the (squared) spin non flip term. As anticipated, 
the quality of the fit has not been altered. 

Several comments are in order.

(a) The $\IP$ contribution to the spin-flip amplitude $h(s,t)$ is 
considerably smaller than to the spin-non-flip term $g(s,t)$ (about 5\%, 
roughly, at $\sqrt{s}=$ 20 GeV and changing little until 500 GeV). This 
value is compatible with the analysis performed in \cite{buttimore99} 
where use was made of the relative amplitude $r_5=m\phi_5/(\sqrt{-t}\;{\rm 
Im}\phi_{+})$. In our case, ${\rm Im}\; r_5=-0.054$ at $\sqrt{s}=$ 500 
GeV. We should mention that we find {\it natural} that the spin-flip part 
of the amplitude should be a small fraction of the spin-non-flip but, 
again, we have no real reason for this.
(b) Contrary to the discussion made in Ref. \cite{hinotani79}, the 
(small $|t|$) slope of the spin-flip amplitude $\beta_1=
4.74\;{\rm GeV}^{-2}$ appears to be not very much different from the 
effective slope of the Pomeron in spin-non-flip amplitude (see the 
Appendix \ref{appe:spin}). The present parametrization, however, is 
considerably more elaborate and the set of data corresponds to 
higher energies so the results of these papers may not be directly 
comparable.
(c) Our result [i.e., $h(s,t)$] cannot be extended to $|t|$ values 
much higher than few ${\rm GeV}^2$ 
because the spin-non-flip amplitude utilized is valid at the Born level
\cite{desgrolard00} and its description in the region after the dip ($|t|> 
1.5\;{\rm GeV}^2$) is not very good. To extend our considerations to 
higher $|t|$, it would be necessary to adopt the more sophisticated 
eikonalized version. Anyway, the $t$ region of interest for RHIC is up to 
$1.5\;{\rm GeV}^2$ \cite{guryn00} so we can confine our analysis to the 
not too high $|t|$ region. This point may have to be reconsidered in the 
future along with more detailed analysis.
(d) The spin-non-flip amplitude in Ref. \cite{desgrolard00} 
fits $d\sigma/dt$ without the spin contribution and that description 
is not spoiled by the presence of $h(s,t)$ on this work since it is 
very smaller than $g(s,t)$.

\subsection{The kinematical zero of the spin-flip amplitude}
\label{subsec:kine}

The original work about diffraction and polarization data at $\pi p$ 
scattering calculated the reduced spin-flip amplitude removing the 
kinematical zero by means of a $\sin\theta$ instead of $\sqrt{-t}$ 
\cite{predazzi67}. Since the energies available at that time were much 
lower, it would be difficult to see differences between the two 
approaches. But the data presently available and the coming in operation 
of RHIC at even higher energies raises the question of what would happen 
if the factor $\sin\theta$ was used. In principle, the use of 
relativistically invariant variables used earlier seems more appropriate 
but we feel that the answer can only come from experiments.

To answer this question we adopt in this section the spin-flip 
amplitude

\begin{eqnarray}
h(s,t)&=&a^{sf}(s,t) = (i\gamma_1+\delta_1)\sin\theta\;
\tilde{s}^{\alpha^{sf}(t)}e^{\beta_1 t}\Theta(0.5-|t|)   
\nonumber\\
&+& (i\gamma_2+\delta_2)\sin\theta\;
\tilde{s}^{\alpha^{sf}(t)}e^{\beta_2 t}\Theta(|t|-0.5) .
\label{spinamplsin}
\end{eqnarray}
We use the same procedure of the fit with Eq. (\ref{spinampl2}), 
that is, we fit the data at $\sqrt{s}=13.8$, 16.8 and 23.8 GeV 
assuming $\alpha^{sf}(t)=\alpha_{\IP}(t)$. Then we check the quality 
of the fit with the description of the polarization data at 19.4 GeV. 
The values of the parameters obtained with Eq. (\ref{spinamplsin}) 
are shown in Table \ref{tab6paramsin}.

\begin{figure}
\includegraphics[width=9cm,height=7cm]{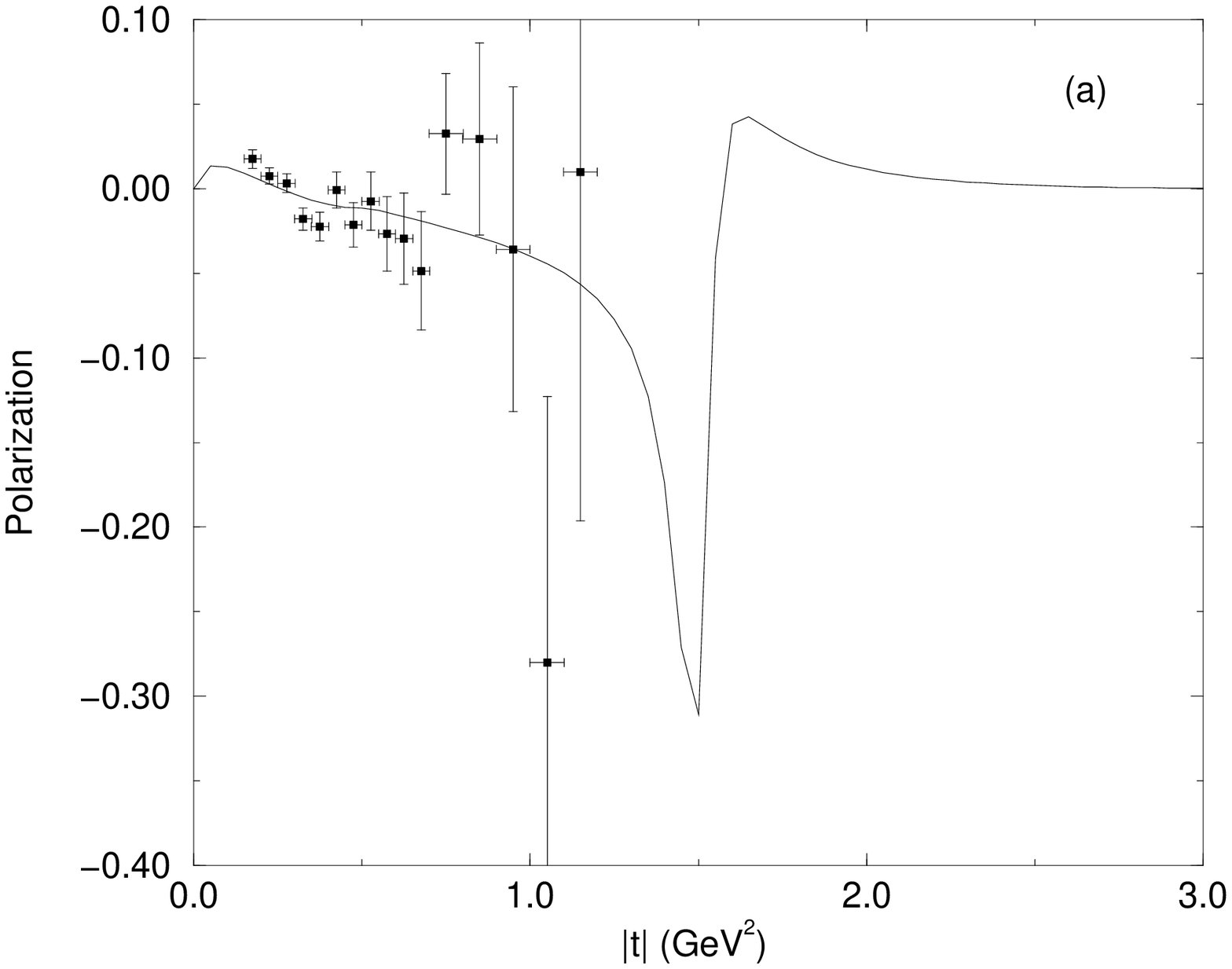}
\includegraphics[width=9cm,height=7cm]{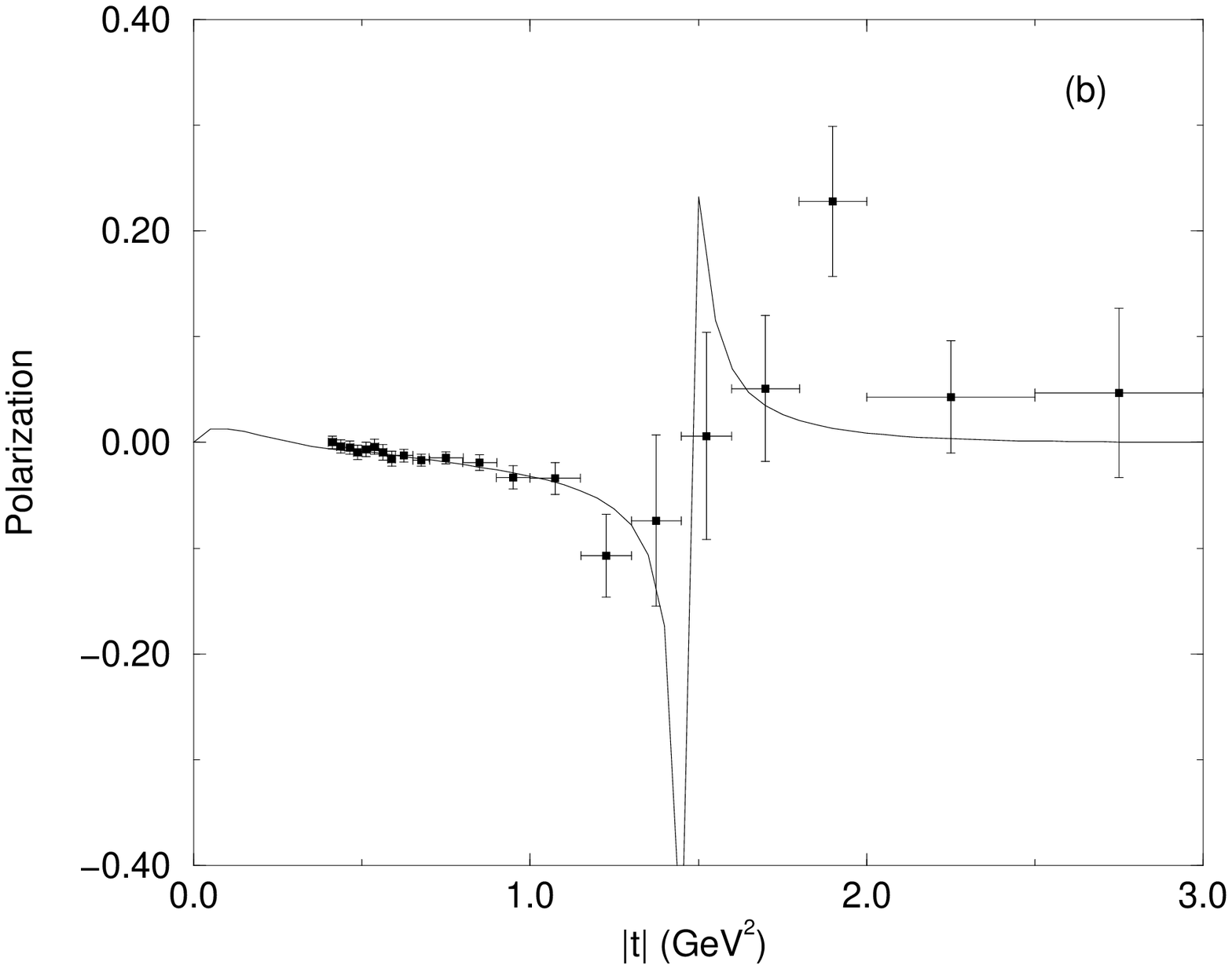}
\includegraphics[width=9cm,height=7cm]{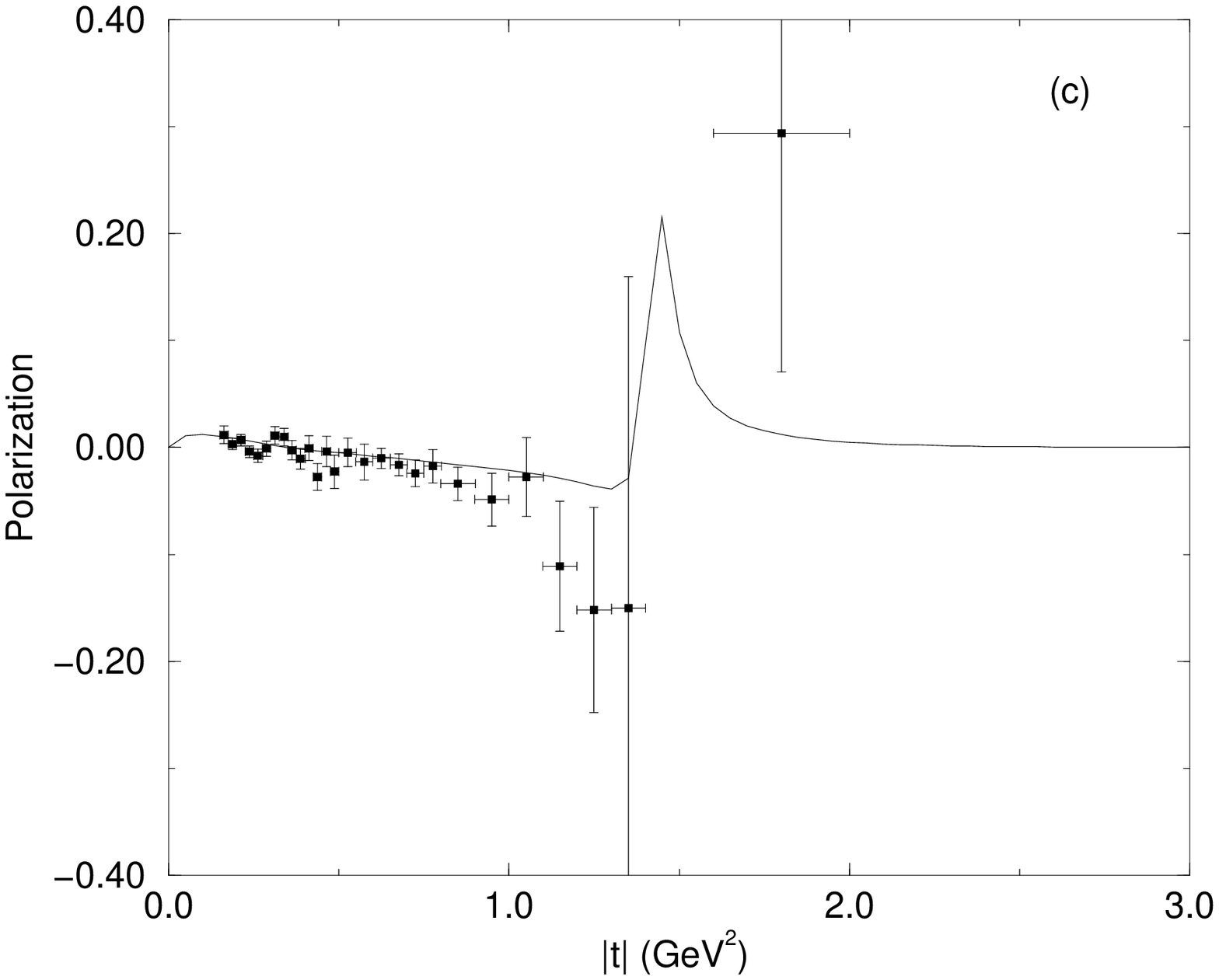}
\caption{\label{figpol6paramsin}Results from fitting polarization data 
at (a) 13.8 GeV, (b) 16.8 GeV, and (c) 23.8 GeV (see Table
\protect\ref{tab6paramsin}).}
\end{figure}

\begin{figure}
\includegraphics[width=10cm,height=8cm]{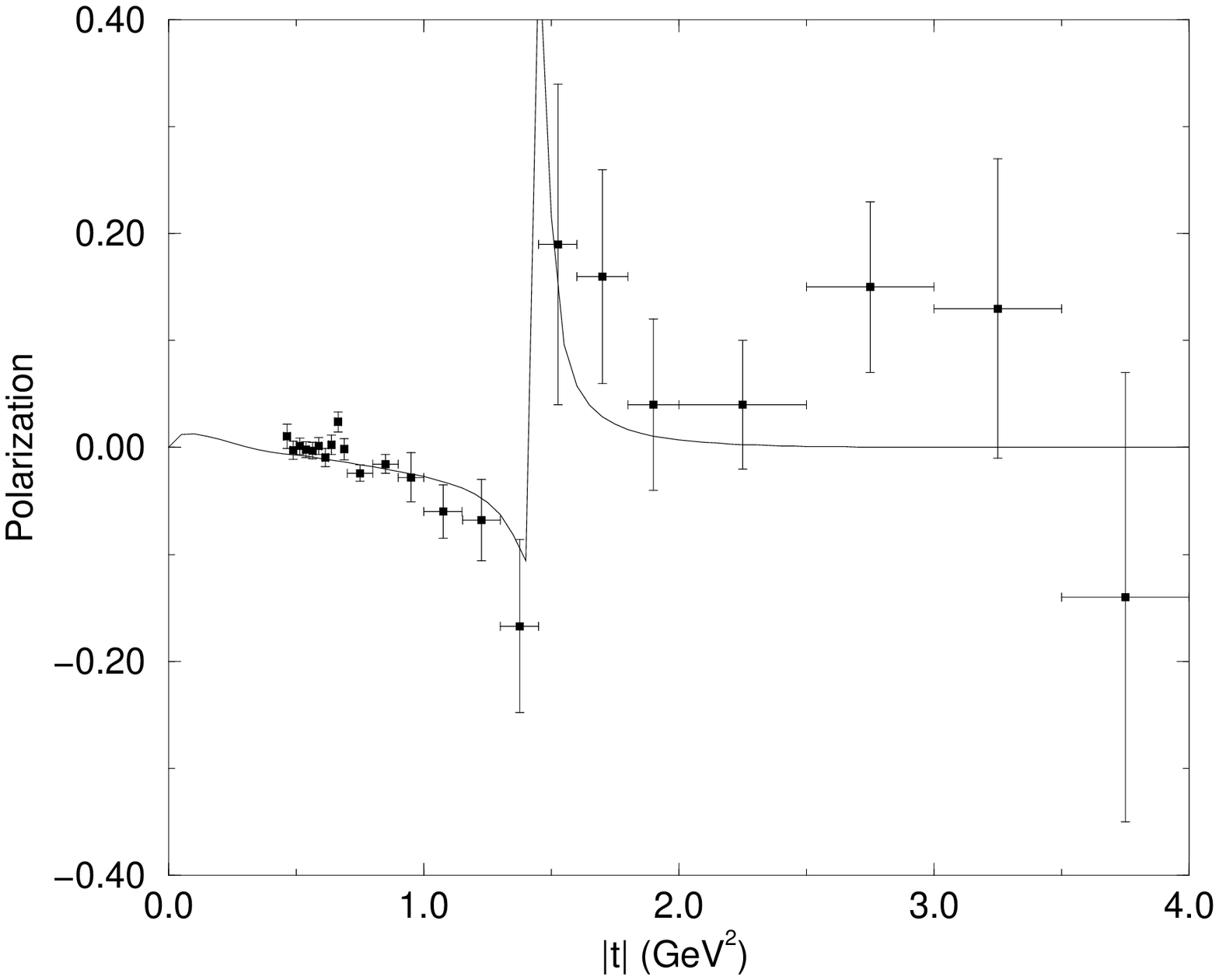}
\caption{\label{figpol19pt4sin}The prediction for the polarization at 
19.4 GeV (not used in the fit) compared with the experimental data at 
that energy (parameters from Table \protect\ref{tab6paramsin}).}
\end{figure}

\begin{figure}
\includegraphics[width=12cm,height=10cm]{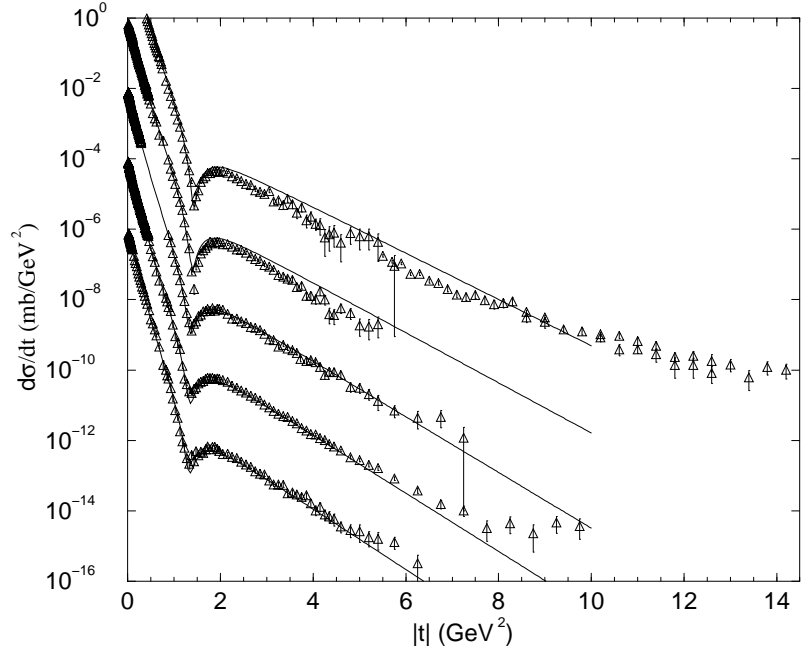}
\caption{\label{figdsdtsin}The differential cross section obtained in this 
work taking into account the spin-flip amplitude through Eq. 
(\protect\ref{spinamplsin}). About the data see Fig. 
\protect\ref{figdsdt}.}
\end{figure}

Following the same procedure of the previous section, we present 
in Fig. \ref{figpol6paramsin} the result of the fitting 
while Fig. \ref{figpol19pt4sin} shows the polarization at 
$\sqrt{s}=19.4$ GeV and Fig. \ref{figdsdtsin} presents $d\sigma/dt$. 
Two differences in the present results and in those of the previous 
analysis are worth being emphasized. First, the ratio 
of the reduced spin-flip [$h(s,t)/\sin\theta$] to the spin-non-flip 
amplitude is much bigger (above 90\%) and this big ratio was noticed at 
lower energies too \cite{hinotani79}. Second, the slope $\beta_1=6.25
\;{\rm GeV}^2$ is higher than the previous result 
(Table \ref{tab6param}) so the reduced spin-flip amplitude is 
steeper on this case. 

Although Eqs. (\ref{spinampl2}) and (\ref{spinamplsin}) 
present very different values for the parameters $\beta_1,\;\delta_1$, 
and $\gamma_1$, Figs. \ref{figpol6param}-\ref{figdsdtsin} show 
that the descriptions of the data are similar and the fits have a 
comparable $\chi^2$. To understand the consequences of using one 
parametrization or the other it is necessary to remember that 
$\sin\theta\propto\sqrt{-t/s}$. So, to all effects, Eq. 
(\ref{spinamplsin}) does not have the $s$ dependence of the Pomeron. 
Differently stated, the parametrization (\ref{spinamplsin}), ultimately, 
is not diffractive. The comparison between the two parametrizations used 
for $h(s, t)$, therefore, Eqs. (\ref{spinampl2}) and 
(\ref{spinamplsin}) goes to the root of the problem of whether or not 
diffraction contributed to spin-flip. This difference will have relevant 
consequences at higher energies and the $pp2pp$ experiment at RHIC 
\cite{guryn00} will be able to discriminate between these two scenarios.

\subsection{Predictions to higher energies}
\label{subsec:predi}

\begin{figure}
\includegraphics[width=10cm,height=8cm]{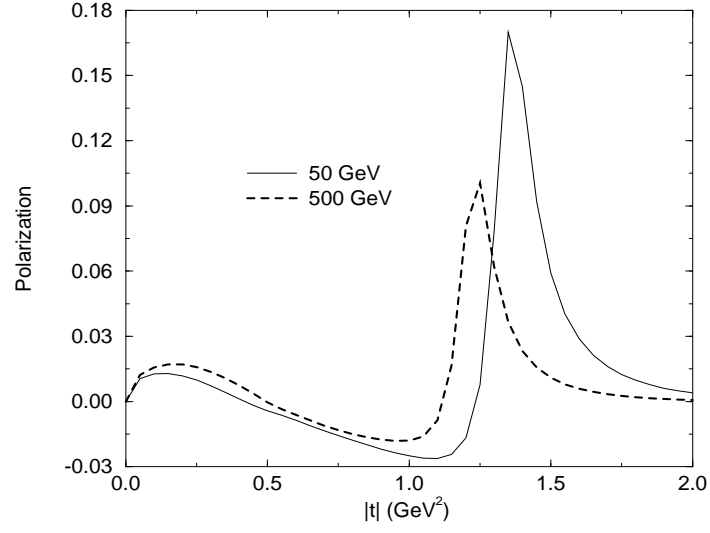}
\caption{\label{figpol50500}The polarization predictions for 
$\protect\sqrt{s}=50$ and 500 GeV
with the parameters of Table \protect\ref{tab6param}.}
\end{figure}

The $pp2pp$ experiment at RHIC will provide information about polarization 
in $pp$ scattering at energies between 50 and 500 GeV \cite{guryn00}. 
That means a big increase in the amount of information about the spin 
content of the proton so it is important to see what we can say about the 
polarization in that energy range based on the information we 
have derived from the data available at lower energies. To this aim, we 
calculate the polarization at $\sqrt{s}=50$ and 500 GeV with 
Eq. (\ref{spinampl2}) plotting the result in Fig. \ref{figpol50500}. 
We can see that the polarization decreases in magnitude with increasing 
energy but it is still sizeable at 500 GeV. Equation (\ref{spinampl2}) 
produces a peak of about 10\% of positive polarization around 
$-t=1.25\;{\rm GeV}^2$ and $\sqrt{s}=500$ GeV as compared with a 
(positive) polarization of over 15\% at $\sqrt{s}=50$ GeV.

\begin{figure}
\includegraphics[width=10cm,height=8cm]{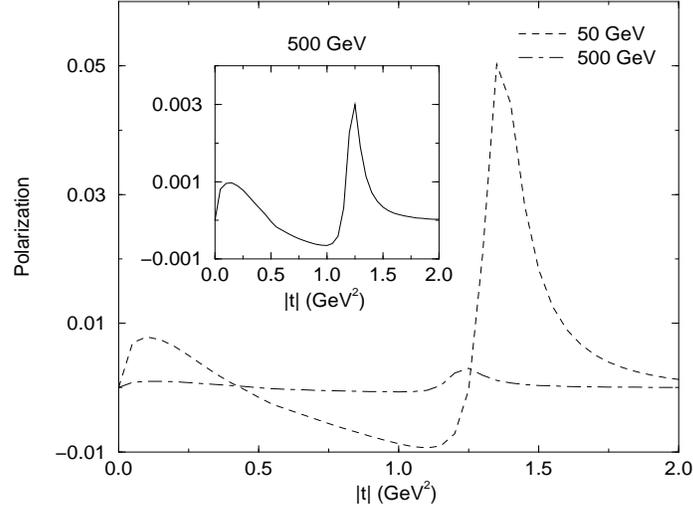}
\caption{\label{figpol50500sin}The polarization predictions for 50 and 500 
GeV with a detailed view of the data at 500 GeV in the inset 
(parameters from Table \protect\ref{tab6paramsin}).}
\end{figure}

Now we compare this result with the prediction using Eq. 
(\ref{spinamplsin}) (see Fig. \ref{figpol50500sin}). The polarization 
calculated with the factor $\sin\theta$ is much smaller (about 5\% at 
$\sqrt{s}=50$ GeV) but becomes essentially zero at $\sqrt{s}=500$ GeV.
As already mentioned the factor $1/\sqrt{s}$ hidden inside the sine 
function is important to separate the predictions at high energies.

We have also checked the possibility of a different intercept for the 
trajectory $\alpha^{sf}(t)$. In this case Eq. (\ref{alfsf}) is 
modified and $\alpha^{sf}(0)$ becomes a new parameter to be fitted 
together with the other six parameters. The values obtained (with a 
similar $\chi^2/N_{\rm DF}$) for them were quite absurd, 
including a negative intercept $\alpha^{sf}(0)$ when the 
factor $\sqrt{-t}/m_p$ is used, 
and the polarization predictions at RHIC energies are so small that 
it would be essentially impossible to detect those values 
experimentally. Since there is no improvement of the fit with the 
increased number of parameters we discard this solution.

\section{Conclusions}
\label{sec:concl}

We analyzed the spin-flip amplitude and removed its kinematical zero 
to study the Pomeron contribution to the spin as suggested long time 
ago \cite{predazzi67,hinotani79}. We made use of two hypothesis to 
remove the zero, $\sqrt{-t}/m_p$ and $\sin\theta$, where the first 
is more convenient at high energies since it is relativistic invariant 
while the second was used in the original work about diffraction 
in $pp$ spin-flip amplitudes \cite{hinotani79}. The differences 
resulting from the application of Eqs. (\ref{spinampl2}) or 
(\ref{spinamplsin}) are quite large although the descriptions of the 
data between $\sqrt{s}=$ 13 and 24 GeV are very similar. The 
fraction of $h(s,t)$ to $g(s,t)$ is small if $\sqrt{-t}/m_p$ is used 
(around 5\%) but it becomes big when $\sin\theta$ is utilized (more than 
90\%). Although, as already mentioned, we do not have a strict reason to 
prefer one over the other, the first solution is much more in line with
the traditional analyses.
Also, the slope at small $|t|$ changes when $\sqrt{-t}/m_p$ is 
substituted by $\sin\theta$, showing that $h(s,t)$ becomes steeper 
with the second option. At the same time, the slope $\beta_2$ is 
practically the same in both cases showing that there is a modification 
on $h(s,t)$ around $-t=0.5\;{\rm GeV}^2$, which confirms the necessity 
of two slopes in the spin-flip amplitude as it was assumed. The 
extrapolations shown in Figs. \ref{figpol50500} and \ref{figpol50500sin} 
suggest that the data from RHIC will be crucial to understand and choose 
the best form to describe the spin-flip amplitude.

In summary, the prediction is very straightforward: a positive 
polarization of the order of 10\% is predicted at RHIC energies of 
$\sqrt{s}=500$ GeV if the Pomeron (diffraction) contributes to the 
spin-flip amplitude as lower energy data tend to suggest. A vanishing 
contribution is otherwise expected.

It would be possible to improve the description of the data with more 
elaborated forms for both spin-non-flip and spin-flip amplitudes (by 
eikonalizing them, for example, Ref. \cite{desgrolard00}, by introducing 
subasymptotic effects in the spin-flip part by secondary Reggeons and so 
on) but we expect that the main conclusions of our analysis would remain 
the same in the region of small $t$ (where the Born amplitudes work well) 
and high energies (where the Pomeron dominates). 
A more detailed analysis to separate the contributions of the exchanges of 
two and three gluon ladders would be more gratifying from the theoretical 
point of view as well as a better study of the break in the diffractive 
slopes. All these points will, hopefully, be reconsidered in the future.

\begin{acknowledgments}
Several discussions with Professor E. Martynov and Professor 
M. Giffon and the comments and suggestions of Professor O. V. Selyugin 
are gratefully acknowledged. Work supported in part by the MURST of Italy. 
One of us (A.F.M.) would like to thank the Department of Theoretical 
Physics of the University of Torino for its hospitality and the 
FAPESP of Brazil for its financial support.
\end{acknowledgments}

\appendix*

\section{\label{appe:spin}The spin-non-flip amplitude}

The spin-non-flip amplitude utilized in this work is
\begin{equation}
a^{nf}(s,t)\equiv a_{+}(s,t)-a_{-}(s,t) ,
\end{equation}
where
\begin{equation}
a_{+}(s,t)=a_{\IP}(s,t)+a_f(s,t) 
\end{equation}
and
\begin{equation}
a_{-}(s,t)=a_{O}(s,t)+a_{\omega}(s,t).
\end{equation}

The expressions for the two Reggeons used in Ref. \cite{desgrolard00} 
are
\begin{equation}
a_R(s,t)=a_R\tilde{s}^{\alpha_R(t)}e^{b_R t}
\end{equation}
and
\begin{equation}
\alpha_R(t)=\alpha_R(0)+\alpha_R't\; (R=f\; {\rm and} \; \omega)
\end{equation}
with $a_f (a_{\omega})$ real (imaginary). 

For the Pomeron, the spin-non-flip amplitude is
\begin{equation}
a_{\IP}^{(D)}(s,t)=a_{\IP}\tilde{s}^{\alpha_{\IP}(t)}
[e^{b_{\IP}[\alpha_{\IP}(t)-1]}(b_{\IP}+\ln\tilde{s})+d_{\IP}\ln\tilde{s}] 
\end{equation}
while for the Odderon we use
\begin{equation}
a_{O}(s,t)=[1-\exp(\gamma t)]a_O\tilde{s}^{\alpha_O(t)}
[e^{b_O(\alpha_O(t)-1)}(b_O+\ln\tilde{s})+d_O\ln\tilde{s}],
\end{equation}
where again $a_{\IP} (a_O)$ real (imaginary) and we have 
utilized $\alpha_i(t)=\alpha_i(0)+\alpha_i't$, where $i=\IP ,O$.

Our definition for the amplitude follows \cite{desgrolard00} so 
that
\begin{eqnarray}
\sigma_t={4\pi\over s}{\rm Im}\{ a^{nf}(s,t=0)\},
\label{sctot} \\
{d\sigma\over dt}={\pi\over s^2}[|a^{nf}(s,t)|^2+2|a^{sf}(s,t)|^2].
\label{scdif}
\end{eqnarray}

In this work we retain the same parameters for the spin-non-flip
amplitude as in Ref. \cite{desgrolard00} and we keep them fixed while 
fitting the parameters of the spin-flip 
amplitude. We utilize the dipole model at the Born level since 
most of the polarization data is contained in the $t$ domain 
corresponding to the region before the dip in $d\sigma/dt$ (well 
described without eikonalization). The values of the parameters of 
the spin-non-flip amplitude \cite{desgrolard00} are shown in 
Table \ref{tabsnfpar}.

To calculate the polarization we utilized the form
\begin{equation}
P=2
{{\rm Im}\{a^{nf}(s,t)[a^{sf}(s,t)]^{\star}\}\over 
|a^{nf}(s,t)|^2+2|a^{sf}(s,t)|^2} ,
\end{equation}
where the star on the numerator means complex conjugate.

\begin{table}[hbt]
\caption{\label{tab6param}Values of the parameters obtained from fitting 
polarization data at $\protect\sqrt{s}=13.8$, 16.8, and 23.8 GeV with 
Eqs. (\protect\ref{spinampl2}) and (\protect\ref{alfsf}). The 
$\chi^2/N_{\rm DF}$ is 1.1.}
\begin{ruledtabular}
\begin{tabular}{cccc}
i & $\gamma_i$ & $\delta_i$ & $\beta_i\;({\rm GeV}^{-2})$\\
\hline
1 & $1.35\times 10^{-1}$ & $2.64\times 10^{-1}$ & 4.74\\
2 & $2.55\times 10^{-2}$ & $5.38\times 10^{-2}$ & 2.29\\
\end{tabular}
\end{ruledtabular}
\end{table}

\begin{table}[hbt]
\caption{\label{tab6paramsin}Results from fitting polarization data at 
$\protect\sqrt{s}=13.8$, 16.8, and 23.8 GeV with
Eqs. (\protect\ref{alfsf}) and (\protect\ref{spinamplsin}). 
The $\chi^2/N_{\rm DF}$ is 1.1.}
\begin{ruledtabular}
\begin{tabular}{cccc}
i & $\gamma_i$ & $\delta_i$ & $\beta_i\;({\rm GeV}^{-2})$\\
\hline
1 & 2.55 & 4.80 & 6.25\\
2 & 0.18 & 0.45 & 2.30\\ 
\end{tabular}
\end{ruledtabular}
\end{table}

\begin{table}[htb]
\caption{\label{tabsnfpar}Parameters of the dipole model at the Born level 
\protect\cite{desgrolard00} with $i=\IP ,O,f,\omega$.}
\begin{ruledtabular}
\begin{tabular}{ccccc}
 & Pomeron & Odderon & $f$ Reggeon & $\omega$ Reggeon \\ 
\hline
$\alpha_i(0)$ & 1.071 & 1.0 & 0.72 & 0.46 \\
$\alpha_i'\;({\rm GeV}^{-2})$ & 0.28  & 0.12 & 0.50 & 0.50 \\
$a_i$ & -0.066 & 0.100 & -14.0 & 9.0 \\
$b_i$ & 14.56 & 28.10 & 1.64 ${\rm GeV}^{-2}$ & 0.38 ${\rm GeV}^{-2}$ \\
$d_i$ & 0.07 & -0.06 & - & - \\ 
$\gamma\;({\rm GeV}^{-2})$ & - & 1.56 & - & - \\
\end{tabular}
\end{ruledtabular}
\end{table}

\end{document}